\documentclass[submission, Phys]{SciPost}
\hypersetup{
    colorlinks,
    linkcolor={red!50!black},
    citecolor={blue!50!black},
    urlcolor={blue!80!black}
}

\usepackage[bitstream-charter]{mathdesign}
\DeclareSymbolFont{usualmathcal}{OMS}{cmsy}{m}{n}
\DeclareSymbolFontAlphabet{\mathcal}{usualmathcal}
\def\KD{K\"{a}hler-Dirac }
\newcommand{\cK}{{\cal K}}
\def\Phib{\overline{\Phi}}
\def\Psib{\overline{\Psi}}
\def\phib{\overline{\phi}}

\begin{document}

\begin{center}{\Large \textbf{Lattice Regularization of Reduced \KD Fermions and Connections to
Chiral Fermions
}}\end{center}

\begin{center}
Simon Catterall
%\textsuperscript{$\star$}
\end{center}

% TODO: write all affiliations here.
% Format: institute, city, country
\begin{center}
{\bf } Department of Physics, Syracuse University, Syracuse, NY 13244, USA 

% TODO: provide email address of corresponding author
%${}^\star$ 
{\small \sf smcatter@syr.edu}
\end{center}

\begin{center}
\today
\end{center}

\section*{Abstract}
{\bf
% TODO: write your abstract here.
We show how a path integral for reduced \KD fermions suffers from a phase ambiguity
associated with the fermion measure that is an analog
of the measure problem seen for chiral fermions. However, unlike the case of chiral
symmetry, a doubler free lattice action exists which is invariant
under the corresponding onsite symmetry. This allows for a clear diagnosis
and solution to the problem using mirror fermions resulting in a unique gauge invariant measure.
By introducing an appropriate set of Yukawa interactions which are consistent with 't Hooft
anomaly cancellation we conjecture the mirrors can be decoupled from low energy physics.
Moreover,
the minimal such \KD mirror model yields a light sector
which corresponds, in the flat space continuum limit, to the Pati-Salam
GUT model.
}

\vspace{10pt}
\noindent\rule{\textwidth}{1pt}
\tableofcontents\thispagestyle{fancy}
\noindent\rule{\textwidth}{1pt}
\vspace{10pt}

\section{Chiral symmetry, \KD fermions and the lattice}
Efforts to give a non-perturbative definition of a chiral gauge theory
go back to the earliest days of lattice field theory. Unfortunately the Nielsen-Ninomiya
theorem puts strong constraints on attempts to construct theories with chiral lattice fermions \cite{Nielsen:1981hk}. In essence
it says that any lattice theory with an exact $\gamma_5$ symmetry will
necessarily contain equal numbers of left and right handed fields. This is naturally understood
on the basis of anomalies - the exact chiral symmetry of the lattice theory 
should naively survive quantization and hence cannot generate the required
chiral anomaly in the continuum limit. The lattice
achieves this by automatically generating a vector-like theory via the well known
phenomenon of fermion doubling.

The discovery of domain wall and overlap fermions \cite{Kaplan:1992bt,PhysRevLett.71.3251,Shamir:1993zy,NEUBERGER1998141,LUSCHER1998342} 
led to doubler free lattice actions
that are invariant under a modified and mildly non-local
lattice chiral symmetry - see \cite{Clancy:2023ino} for recent work on these
constructions. The nature of this lattice symmetry is such that it 
is capable of producing the correct anomaly for the global
chiral symmetry of Dirac fields. 
However, because the associated chiral projector is non-local
and depends on the background gauge field the associated path integral for
a Weyl field picks up a phase depending on the 
gauge field. Furthermore this phase is not uniquely determined -- it can be changed by
performing a unitary rotation of the orthonormal bases for the chiral
fermions -  this is the well known measure problem \cite{Poppitz:2010at} which
renders the theory ill-defined unless a prescription is given to determine this
phase. 
While such a prescription can be shown to exist for (anomaly free) abelian theories 
there is currently no non-perturbative construction for
anomaly free non-abelian theories \cite{Martin_Luscher_2000}.

One strategy to get around this measure problem involves building a mirror model which starts
from the target chiral theory and adds mirror
fermions with opposite chirality. The resultant theory, being
vector-like, possesses a well defined measure. One then attempts to give masses
to the mirror fermions without disturbing the symmetries and dynamics of the target light chiral
sector. Many efforts 
to realize such mirror models using different types of lattice fermion have been tried - see \cite{Eichten:1985ft, Poppitz:2009gt}
for a review and a discussion of the most recent and best motivated effort using 
overlap fermions or domain wall fermions \cite{Grabowska:2015qpk}.
However, they have so far not yielded positive results~\footnote{The exception to this seems to be abelian theories in two dimensions where appropriate six fermion interactions have been shown to successfully decouple the mirror states leaving a low energy chiral theory\cite{Wang:2013yta,Wang:2018ugf,Zeng:2022grc}. See also the recent construction in \cite{Berkowitz:2023pnz}.}

In this paper we describe an analog of this problem
in the case where the Dirac equation is replaced by the \KD equation \cite{kahler}. The relevant
symmetry is no longer generated by $\gamma_5$ but by a new operator $\Gamma$ that is
unique to \KD fermions \cite{Butt:2021brl}. The analog of a chiral fermion is a so-called
{\it reduced} \KD (RKD) field.  
However, the crucial difference between chiral and RKD fermions 
is that a doubler free 
lattice action can be constructed which is invariant
under precisely the same $U(1)$ symmetry as the continuum theory. Indeed, this 
lattice theory can reproduce {\it exactly} the analog of the axial anomaly seen for
chiral fermions. The anomaly
arises from the variation of the fermion measure just as in the continuum and captures
the non-trivial topology of the background space \cite{Catterall:2018lkj,Catterall:2018dns}.
It is thus a mixed $U(1)$-gravitational anomaly.
These features
allow for the construction of a lattice action for RDK fermions 
where the measure problem can be made explicit and solved using
a mirror fermion construction. The additional issue of anomaly cancellation can also
be stated precisely and solved even on the lattice. 

In addition, we propose a set Yukawa interactions that respect the 't Hooft anomaly
cancellation conditions and should be capable of driving the mirrors into
a symmetric gapped phase without breaking symmetries of the light sector.

Finally, we point out that the minimal, anomaly free \KD mirror
model yields a theory with the global symmetries and matter representations
of the Pati-Salam GUT \cite{Pati} in the naive flat space
continuum limit suggesting that lattice \KD constructions are capable of regulating at least a class of chiral gauge theories. In our paper we will mostly
focus on four Euclidean dimensions but our analysis
can be easily extended to any arbitrary (even) dimension.

\section{\KD fermions on and off the lattice}
The $D$-dimensional continuum \KD action can be written as
\begin{equation}
    S=\left[\Phib, \left(K+m\right)\Phi\right]\label{KDaction}
\end{equation}
where $K=d-d^\dagger$ is a natural square root of the Laplacian and hence a candidate
for a fermion operator. The corresponding fermion field
$\Phi$ comprises a set of Grassman valued antisymmetric tensor fields ($p$-forms)
\begin{equation}
\Phi=(\phi,\phi_\mu,\phi_{\mu_1\mu_2},\ldots ,\phi_{\mu_1\ldots\mu_p},\ldots,\phi_{\mu_1\cdots\mu_D})
\end{equation}
The action of $d$ and $d^\dagger$ on these component forms is given by \cite{Banks:1982iq}
\begin{equation}
d\Phi=\left(0,\partial_\mu\phi, \partial_{\mu_1}\phi_{\mu_2}-\partial_{\mu_2}\phi_{\mu_1},\ldots ,\sum_{perms\, \pi}\left(-1\right)^\pi\partial_{\mu_1}\phi_{\mu_2\ldots\mu_D}\right)\end{equation}
\begin{equation}
-d^\dagger\Phi=\left(\phi^\nu,\phi^\nu_\mu,\ldots ,\phi^\nu_{\mu_1,\ldots ,\mu_{D-1}},0\right)_{;\nu}.\end{equation}
and the inner product between two \KD fields $A$ and $B$ is defined as
\begin{equation}
\left[A,B\right]=\int d^Dx\,\sqrt{g}\sum_p \frac{1}{p!}a^{{\mu_1}\ldots{\mu_D}}b_{{\mu_1}\ldots{\mu_D}}.\end{equation}
In dimension $D$ the field $\Phi$ contains $2^D$ independent components. In a flat space of
even dimensions the
\KD equation describes the propagation of $2^{D/2}$ degenerate Dirac fields. To see this
first construct matrices $\Psi$ and $\Psib$ from elements of the Clifford algebra
of $D$-dimensional (Euclidean) Dirac matrices using the p-form components of $\Phi$ as coefficents:
\begin{align}
    \Psi&=\sum_p \phi_{\mu_1\ldots\mu_p}\gamma^{\mu_1}\cdots\gamma^{\mu_p}\nonumber\\
    \Psib&=\sum_p \phib_{\mu_1\ldots\mu_p}\left(\gamma^{\mu_1}\cdots\gamma^{\mu_p}\right)^\dagger\label{expansion}
 \end{align}
It can be easily shown that the original \KD action given in eqn.~\ref{KDaction} can then be written
as
\begin{equation}
S=\int d^Dx\,\sum_{a=1}^{2^{D/2}}\Psib^a \left(\gamma^\mu\partial_\mu+m\right)\Psi^a\label{KDmatrix}
\end{equation}
where the index $a$ labels columns of the matrix $\Psi$ each of which corresponds to a Dirac spinor~\footnote{In odd dimensions the mapping to spinors is more subtle -- the matrix representation of the $D=2n+1$ dimensional \KD theory
requires employing gamma matrices of size $2^{n+1}$ \cite{Catterall:2022ukg}. The degrees of freedom carried by
the resultant matrix fermion are then reduced by a factor of two
using the projectors described in section~\ref{reduced}.}. This equivalence
between a \KD field and a multiplet of degenerate Dirac fields is no longer true
in a curved space as the gravitational coupling of tensors is very different from
spinors.
Indeed this observation lies at the heart of why there are new
anomalies of \KD fermions that are different from the usual anomalies of Dirac fermions.
In particular \KD fermions are allowed on non spin manifolds and do
not require the additional structure of frames and spin connections. One obvious
manifestation of this difference is the observation that
the \KD equation
possesses zero modes even on spaces with positive curvature such as the sphere $S^D$.

The flat space action in eqn.~\ref{KDmatrix} is invariant under both (Euclidean) $SO(4)$ Lorentz symmetry
and an $SU(2^{D/2})$ flavor symmetry. 
\begin{equation}
    \Psi\to L\Psi F^\dagger
\end{equation}
However the original \KD theory is invariant
only under a diagonal subgroup of these symmetries corresponding to the matrix transformation
\begin{equation}
    \Psi\to L\Psi L^\dagger
\end{equation}
This transformation makes it clear how the Lorentz transformation properties
of a set of spinors can be compatible with a collection of tensors~\footnote{\KD fermions are most naturally formulated in Euclidean signature \cite{Banks:1982iq}.
Extensions to Minkowski
signature are possible but typically lead to non-compact flavor symmetries.}. One simply employs the relation
$L\gamma_\mu L^\dagger=\Lambda^\nu_\mu \gamma_\nu$ where $\Lambda^\nu_\mu$ is
the usual boost matrix for vectors together with the expansion of $\Psi$ given in
eqn.~\ref{expansion}.

We will be particularly interested in an operator $\Gamma$ 
whose action on $p$-form fields is given
by 
\begin{equation}
    \Gamma\phi_{\mu_1\ldots \mu_p}=\left(-1\right)^p\phi_{\mu_1\ldots\mu_p}
\end{equation}
It is easy to see that $\Gamma$ anticommutes with $K$. This allows
us to construct a $U(1)$ transformation 
\begin{align}
    \Phi&\to e^{i\alpha\Gamma}\Phi\nonumber\\
    \Phib&\to \Phib e^{i\alpha\Gamma}
\label{u1}
\end{align}
This is a symmetry of the massless action but we will show that this is not a symmetry of the theory
in curved space due to an anomaly. In the matrix representation it can be realized as
\begin{equation}
    \Psi\stackrel{\Gamma}{\to}\gamma^5\Psi\gamma^5
\end{equation}
and in this context it is sometimes referred to as a twisted chiral symmetry. 

One of the prime advantages of the \KD equation over the
Dirac equation is that it can be formulated
on a discrete approximation to a continuum space such as a triangulation \cite{Rabin:1981qj}. The $p$-form
fields are to be replaced by $p$-cochains -- lattice fields defined on $p$-simplices.
Each $p$-simplex is specified by a list of $(p+1)$ vertex labels $\left[a_0,\ldots a_p\right]$ and the p-simplex field  $\Phi^{(p)}$ is given by the formal sum
\begin{equation}
   \Phi^{(p)}= \sum_{\rm p-simplices} \left[a_0,\ldots, a_p\right] \phi^{(p)}_{\left[a_0,\ldots, a_p\right]}
\end{equation}
Discrete analogs of $d$ and its adjoint $d^\dagger$ exist - the
co-boundary $\bar\delta$ and boundary 
$\delta$ operators  \cite{Rabin:1981qj,Becher:1982ud,Catterall:2018lkj,Catterall:2018dns} with
the action of $\delta$ on a p-simplex field being given by
\begin{equation}
    \delta_p {\Phi}^{(p)}=\sum_{k=0}^{p} \left(-1\right)^k\left[a_0,\ldots\hat{a_k},\ldots ,a_p\right] 
    \phi^{(p-1)}_{\left[a_0,\ldots\hat{a_k},\ldots ,a_p\right]}\label{boundaryop}
\end{equation}
where $\hat{a_k}$ denotes the vertex that is removed to get the $k^{\rm th}$ boundary $(p-1)$-simplex. 
It is easy to see that $\delta^2=(\delta^\dagger)^2=0$ just as for the continuum exterior derivative. Using the abbreviated notation
\begin{equation}
C_p\equiv {\left[a_0,\ldots, a_p\right]}
\end{equation} we can write this as
\begin{equation}
    \delta_p \phi(C_p)=\sum_{C_{p-1}} I(C_p,C_{p-1}) \phi(C_{p-1})\label{incident}
\end{equation}
where $I(C_p,C_{p-1})$ is a $N_p\times N_{p-1}$ incidence matrix whose matrix elements are $+1$ if $C_{p-1}$ is
contained in the boundary of $C_p$ with the correct orientation, $-1$ if it occurs with
opposite orientation and zero otherwise.

Furthermore, a natural analog of the continuum \KD action exists for such a triangulation
\begin{equation}
    S=\left[ \Phib, (\delta-\delta^\dagger+m)\Phi\right]
\end{equation}
where $\Phi=\sum_{p=0}^D \Phi^{(p)}$ is the set of all p-cochains in the triangulation, $\delta=\sum_p \delta_p$
and the inner product now involves a sum over the triangulation of
products of p-cochain elements. Crucially, the number of zero modes of the continuum theory,
being related to the ranks of the homology groups, are reproduced exactly in the discrete
theory. This guarantees that no additional fermionic species arise in the lattice theory as compared to
the continuum.

It is also important
to recognize that the properties of
the operator $\Gamma$ and hence the classical $U(1)$ symmetry survive intact under discretization.
Much of our discussion will focus on this discrete case 
where the \KD operator becomes a finite
dimensional matrix. This is a key difference over efforts to treat chiral fermions on the 
lattice where even the most promising efforts require some degree of non-locality in the lattice
realization of $\gamma_5$ or the fermion operator.

\section{\label{reduced}Reduced \KD fermions and a measure problem}
Using the operator $\Gamma$ we can define
the projectors $P_\pm=\frac{1}{2}(1\pm \Gamma)$ and hence decompose a \KD field
into two independent {\it reduced} fields $\Phi_+=P_+\Phi$ and $\Phi_-=P_-\Phi$ with $\Gamma=+1$ and $\Gamma=-1$ respectively. This is analogous
to the decomposition of a Dirac fermion into left and right handed Weyl fermions. Notice
it can be done in the discrete theory with the continuum $\Gamma$ operator - it does not require
the introduction of a modified, non-local $\hat{\gamma}_5$ operator unlike in
overlap fermion constructions of lattice Weyl fermions.
Using these projectors it is then easy to see that the {\it massless} action breaks into two
independent pieces 
\begin{equation}
    S=\left[\Phib_-, K\Phi_+\right]+\left[\Phib_+, K\Phi_-\right]
\end{equation}
Thus the fields $\Phi_+$ and $\Phi_-$ contain only even or odd p-forms or cochains respectively.
The action for a single {\it reduced} \KD fermion with say $\Gamma=+1$ is gotten
by dropping one of these pieces, for example
\begin{equation}
    S_{\rm RKD}= \left[\Phib_-, {K}\Phi_+\right]=\left[\Phib_-,\cK\Phi_+\right]\quad{\rm with}\;{\cK}=P_-KP_+
\end{equation}

Focusing on the discrete case, it is important
to notice that in even dimensions, the operator $\cK$ generically corresponds to a
rectangular matrix since the number
of even and odd simplices are not equal if the space is not flat.
This can be seen explicitly from
the Euler relation
\begin{equation}
    \chi=N_0-N_1+N_2-N_3+\ldots +N_D=N_+-N_-
\end{equation}
where $N_\pm$ denote the numbers of even and odd dimensional simplices.
Thus, if one tries to integrate out the reduced fermions in a triangulation with non-zero
$\chi$ one immediately encounters a problem since one cannot even define ${\rm det}(\cK)$.
The way to proceed is to use singular value decomposition (SVD) to write 
\begin{equation}
    \cK=U\Sigma V^\dagger
\end{equation}
where $U$ is a $N_-\times N_-$ unitary matrix, $V$ is a $N_+\times N_+$ unitary
matrix and $\Sigma$ is a diagonal rectangular matrix with real, positive semi-definite singular
values $\sigma^i,\,i=1\ldots{\rm min}(N_+,N_-)$. 
The columns of $U$ define an orthonormal basis $\{u^i,i=1\ldots N_-\}$ of the odd cochains 
and can be computed as the eigenvectors of $\cK\cK^\dagger$ while the columns of $V$ define a basis $\{v^i,i=1\ldots N_+\}$ for the even cochains and correspond to the eigenvectors
of $\cK^\dagger\cK$. The singular values $\sigma^i$ can be obtained as the (positive) square roots of the non-zero eigenvalues
of either $\cK\cK^\dagger$ or $\cK^\dagger\cK$.
The matrix $\cK$ maps between the $i$th column vector of $V$ and the corresponding
column vector of $U$.
\begin{equation}
    \cK v^i=\sigma^i u^i,\quad i=1\ldots {\rm min}(N_+,N_-)\label{map}
\end{equation}
For the remaining discussion we will focus on the case of a background space with
spherical topology with $\chi=2$ which corresponds to a natural compactification of flat space
and yields $N_+=N_-+2$.
Using these results one can write down an expression for the partition function of
such a reduced \KD system by setting
\begin{align}
    \Psib_-&=\Phib_- U\nonumber\\
    \Psi_+&=V^\dagger \Phi_+\nonumber\\
    D\Phib_-\,D\Phi_+&=D\Psib_-\,D\Psi_+\left[{\rm det}(U^\dagger){\rm det}(V)\right]
\end{align}
Integrating out the fermions ignoring any zero modes yields
\begin{equation}
    Z={\rm det}(U^\dagger){\rm det}(V)\prod_i^{N_-}\sigma_i
\end{equation}
where $\prod_{i=1}^{N_-}\sigma_i=\sqrt{{\rm det}\,(\cK\cK^\dagger)}$ corresponds to the positive
square root of the full \KD operator omitting zero modes.
Clearly the partition function acquires a phase from the product of the two
determinants. What is worse however is that this phase is not unique.
In fact the matrices $U$ and $V$ are not uniquely
determined by the original fermion operator - any column vector of $V$ can be multiplied
by a phase while preserving the orthonormal character of the basis. Because of the mapping given in eqn.~\ref{map} the corresponding column of $U$ must be multiplied by the same
phase. These phases then cancel between ${\rm det}(U^\dagger)$ and ${\rm det}(V)$ {\it except}
for any phase associated with the two unpaired columns of $V$ that correspond to 
zero modes of $\cK$. 

In fact it should be clear there remains an ambiguity in the phase of
fermion measure for any $\chi\ne 0$.
It is analogous to a similar phase ambiguity
seen in the construction of a path integral for
chiral fermions \cite{Poppitz:2010at}.
In the free theory this extra phase can be held fixed and
causes no real difficulty since it cancels out in any observable.

However, this problem becomes more severe in the situation
where one tries to gauge $U(1)$ symmetry given in eqn.~\ref{u1}. 
To gauge the action one simply replaces the incidence
matrices $I(C_p,C_{p-1})$ by $U(1)$-valued gauge fields $W(C_p,C_{p- 1})$ that transform
according to 
\begin{equation}
    W(C_p,C_{p-1})\to \omega^\dagger(C_p)\, W(C_p,C_{p-1})\,\omega^\dagger(C_{p-1})
    \label{gaugelink}
\end{equation}
where $\omega(C_p)=e^{i\alpha(C_p)\Gamma}=e^{i\alpha(C_p)\left(-1\right)^p}$ varies for each element of the $p$-cochain \cite{Catterall:2022jky}.
In this case
the determinants of the unitary matrices $U$ and $V$ depend on the gauge field $W$ and hence
the partition function acquires a gauge field dependent phase
\begin{equation}
    Z=e^{iA(W)}\sqrt{{\rm det}\,(\cK\cK^\dagger)}
\end{equation}
We can then ask whether the effective action $A(W)$ is gauge invariant. Under a gauge
transformation 
\begin{equation}
{\cal K}\to \omega_-^\dagger{\cal K}\omega_+^\dagger\end{equation}
where we have explicitly labeled the gauge rotation $\omega(C_p)$ by whether is applied to even or odd co-chains.
Under this transformation
the matrix $U\to \omega^\dagger_- U$ and $V\to \omega_+ V$ where we view each $\omega_\pm$ as a diagonal
matrix of phases.
The fermion measure will hence pick up an
additional phase factor 
\begin{equation}{\rm det}(\omega_+){\rm det}(\omega_-)
\end{equation} 
Clearly, the imaginary part of the effective action $A(W)$ is
not even gauge invariant and hence the theory of a single reduced \KD fermion  suffers from
a gauge anomaly which ruins the consistency of the theory. It is completely analogous
to the usual $U(1)$ chiral anomaly of a single Weyl field.~\footnote{The 
omission of the zero modes also introduces a non-local constraint into the fermion measure.}

One can imagine restoring gauge invariance by considering a set of reduced fields with
differing charges $q_a$ under the $U(1)$ symmetry. The determinant factors are
replaced by
\begin{equation}
    {\rm det}(\omega_+^{\sum_a q_a})\times {\rm det}(\omega_-^{\sum_a q_a})
\end{equation}Clearly a set of fields satisfying
$\sum_a q_a=0$ will yield a gauge invariant phase $A(W)$. 

However, even in the case where the gauge
anomaly is cancelled, it still suffers from
the ambiguity discussed earlier -- on the sphere
one can still multiply two columns
of $V$ by arbitrary phases which can now vary with the background gauge field $W$.
This would shift $A(W)\to A(W)+\delta(W)$ for arbitrary $\delta(W)$.
This is the fermion measure problem for reduced \KD fields. Without a prescription
for determining $\delta(W)$ the theory is simply ill-defined. It should be contrasted with the measure problem for
chiral fermions where the phase ambiguity arises both from zero modes and additionally
the non-local nature of the projector $\hat{\gamma}_5$.

To address this larger issue one can instead add a set of mirror fermions with the same
charges and symmetries as the original reduced fermions but with the opposite
eigenvalue of $\Gamma$. For simplicity consider a single reduced field with $\Gamma=+1$
and its mirror with $\Gamma=-1$. The resultant (anomaly free)
fermion operator will correspond to a direct sum of $N_-\times N_+$
and $N_+\times N_-$ matrices for the original and mirror sectors respectively
which can be embedded into a square matrix of size $(N_++N_-)\times (N_++N_-)$.
Because of differences in the dynamics of the light and mirror
fermions, the SVD of this matrix will generically lead to a non-zero 
phase but the latter will now be well defined and gauge invariant - the measure
problem will be resolved.~\footnote{Although the theory will still suffer from a sign
problem.}

Of course if this system
is to reproduce a theory of reduced fermions with fixed $\Gamma$ eigenvalue at low energy one will
need to find a suitable set of mirror interactions which are capable
of gapping all the mirror states without affecting the original fermions. We now turn to this
problem.

\section{The minimal anomaly free mirror model}

The simplest mirror model that one can conceive of 
corresponds to adding a mirror field $\Phi_-$ with $\Gamma=-1$.
The corresponding mirror action takes the form
\begin{equation}
    S_{\rm mirror}=\Phib_+ \hat{\cK}\Phi_-
\end{equation}
where $\hat{\cK}=P_+KP_-=-\cK^\dagger$. The combined system is described by a square matrix
of size $N_++N_-$ corresponding to a full \KD field and 
the measure is now well defined.

Actually, even in this case, the theory suffers from a residual problem which needs
to be addressed. Consider the limit where the $U(1)$ gauge coupling is sent to zero
and focus on the global $U(1)$ symmetry. It is straightforward to see that
this symmetry is in fact anomalous by 
considering the case where the gauge rotation $\omega=e^{i\alpha}$ is taken
to be a constant.
Then, the
change in the phase of $Z$ is just given by
\begin{equation}
    Z\to e^{2i\alpha\left(N_+-N_-\right)}Z=e^{2i\chi\alpha}Z
\end{equation}
where the factor of two in the exponent accounts for the presence of two reduced fermions. 
On a background with the topology of the sphere  we
hence see that the global $U(1)$ symmetry is broken to $Z_4$ \cite{Catterall:2018lkj}. 

Thus the symmetry we were were considering in section~\ref{reduced}
is not $U(1)$ but only $Z_4$ on a curved space with the topology of the sphere. If we
imagine gauging this $Z_4$ the anomaly cancellation condition that
guarantees gauge invariance of the measure for a system of reduced
fermions with $\Gamma=+1$ say is then
\begin{equation}
    \sum_a q_a=0\quad {\rm mod}\; 4
\end{equation}
It is interesting to note that that four massless reduced \KD fermions in four
dimensions can
be decomposed into eight
Dirac or sixteen Majorana fermions in the flat space continuum limit. 
This agrees with the critical numbers of
Weyl fermions needed to cancel the spin-$Z_4$ anomaly in flat
space and also the number of boundary chiral fermions that can be gapped in
 five dimensional topological superconductors \cite{Garcia-Etxebarria:2018ajm,Wang:2018cai}. Indeed this condition, combined
with the dimension dependent counting implied by the decomposition of
a \KD field into spinors,
is capable of
predicting the numbers of fermions needed to cancel a variety of
discrete 't Hooft anomaly for free Majorana fermions in any dimension \cite{Guo:2023rnz,Morimoto:2015lua,Kapustin:2014dxa,Fidkowski:2009dba,Wan:2020ynf}.

Of course if the goal is to construct a theory with a light sector containing
only reduced fermions with say $\Gamma=+1$ we will, in addition,
need to find a mechanism that is capable
of lifting all the mirror states with $\Gamma=-1$ to high energy without
disturbing this light sector. 
A necessary condition for such symmetric mass generation to occur is that 
the mirror sector must be free of all 't Hooft anomalies - here the mixed $Z_4$-gravitational anomaly
just described.
Hence, the minimal mirror model of \KD fermions which is free of
such 't Hooft anomalies comprises four reduced fields with say $\Gamma=+1$ representing the light 
fermions and a mirror sector containing four reduced fields with $\Gamma=-1$ (or vice versa).
These four flavors of fermion in each sector would then inherit a global $SU(4)$ symmetry.

The simplest dynamics that we can add to a model with four flavors of
reduced field that is potentially capable of gapping the mirrors corresponds to coupling the fermions
to a scalar field $\phi$ in the  
six dimensional antisymmetric representation of $SU(4)$. To retain the
$Z_4$ symmetry of the model we take these
scalars to change sign under a $Z_4$ transformation which is reminiscent of the
the action of the spin-$Z_4$ symmetry of chiral fermions \cite{Garcia-Etxebarria:2018ajm,Razamat:2020kyf}.
We can rewrite the reduced mirror fields in terms of a doublet 
\begin{equation}\Lambda_{\rm m}=\left(\begin{array}{c}\Phib_+\\\Phi_-\end{array}\right)
\end{equation}
corresponding to the mirror action 
\begin{equation}
    S_{\rm m}=\sum \Lambda^T_m M_{\rm m}\Lambda_m
\end{equation}
where the mirror fermion operator $M_{\rm m}$ has the form
\begin{equation}
    M^{ab}_{\rm m}=\left(\begin{array}{cc}\lambda_{\rm m}\phi^{ab} &\delta^{ab}\hat{\cK}\\
    -\delta^{ab}\hat{\cK}^T&\lambda_{\rm m}\hat{\phi}^{ab}\end{array}\right)
\end{equation}
where 
$\phi$ and $\hat{\phi}$ are $4N_+\times 4N_+$ and $4N_-\times 4N_-$ matrices of scalars
coupled only to the even cochain and odd cochain fields respectively. In fact
these Yukawa couplings are needed to
soak up the zero modes of $\hat\cK$ that occur in the even co-chain sector that would otherwise
make the partition function vanish \cite{Catterall:2018lkj}.
In fact these zero modes are connected to the appearance of 
the global $U(1)$ anomaly discussed earlier and are implied by the
index theorem
\begin{equation}
    n_+-n_-=\chi
\end{equation}
where $n_\pm$ is the number of zero modes of $K$ with eigenvalue $\Gamma=\pm 1$
on a manifold with Euler character $\chi$. 

To generate a four fermion term one can then
add to the action a term that is quadratic in the scalars, but conservation of $Z_4$ actually
allows
for any term that is even in the number of scalars including, for example, a scalar
kinetic term~\footnote{Previous numerical work in four dimensions suggests such a term
may be needed \cite{Butt:2018nkn}}.
The hope would be that if the mirror sector coupling $\lambda_{\rm m}$ is large enough the
mirror sector can be driven into a symmetric massive phase corresponding to
the presence of a four fermion condensate composed of mirror fermions
while leaving the light sector intact - see the review on symmetric
mass generation and references therein \cite{Wang:2022ucy}. 
Evidence in favor of this scenario can be found
in related simulations of reduced staggered fermions in a variety of dimensions \cite{Ayyar:2014eua,Ayyar:2015lrd,Ayyar:2016lxq,Ayyar:2017qii,Catterall:2015zua,Catterall:2016dzf,Butt:2021koj} and in condensed matter physics \cite{You:2017ltx,You:2014vea,Jian:2019zxu,Wu:2019isp}.

In fact one should add a similar term also in the light sector to keep the light sector
partition function non-zero.
Assembling  the light sector fields into an analogous doublet
\begin{equation}
    \Lambda_{\rm l}=\left(\begin{array}{c}\Phib_-\\\Phi_+\end{array}\right)
\end{equation}
The associated fermion operator for the light sector would take the form
\begin{equation}
    M^{ab}_{\rm l}=\left(\begin{array}{cc}\lambda_{\rm l}\hat{\phi}^{ab}&\delta^{ab}\cK\\
    -\delta^{ab}\cK^T&\lambda_{\rm l}\phi\end{array}\right)
\end{equation}
A small bare $\lambda_{\rm l}$ can soak up the zero modes but should be irrelevant in
the continuum limit with the light sector remaining in a massless symmetric phase.

Notice both reduced 
fermion operators now correspond to square, complex, antisymmetric matrices and hence integration
over the fermions yields Pfaffians for both mirror and light sectors. The phases
carried by the matrices $U$ and $V$ defined in our earlier SVD
analysis are now generated by these Pfaffians. If we
have cancelled all anomalies correctly these Pfaffians are free of 't Hooft anomalies
but in general the model will still suffer from a sign problem.

\section{Connection to the Pati-Salam GUT model}

It is interesting
to ask what is the spinor content of the light sector which contains
four copies of say $\Phi_+$ in flat space.
The answer is intriguing -- in the continuum it is precisely a theory with the global
symmetries and matter representations of the Pati-Salam GUT. 
This is clearly seen if we work in the continuum, use the matrix representation for
the field, and choose a chiral basis for the Dirac gamma matrices 
\begin{equation}
\gamma^\mu=\left(\begin{array}{cc}
0&\sigma_\mu\\
\overline{\sigma}_\mu&0\end{array}\right)
\end{equation}
with $\sigma_\mu=(I,i\sigma_i)$ and $\overline{\sigma}_\mu=(I,-i\sigma_i)$
Thus
\begin{equation}
    \Psi^a_+=\left(\begin{array}{cc}
    \psi^a_R&0\\
    0&\psi^a_L\end{array}\right)
\end{equation}
where the index $a=1\ldots 4$ runs over the four copies needed for
anomaly cancellation and we assume that the fermions
transform in the fundamental representation of a global $SU(4)$ symmetry.
The $2\times 2$ blocks $\psi_L$ and $\psi_R$ are composed of two doublets of Weyl fermions transforming under an ``internal" $SU(2)\times SU(2)$ flavor symmetry in
addition to $SU(4)$. If we imagine continuing back
to Minkowski space this fermion content corresponds to a set of sixteen left-handed
Weyl spinors transforming in the representations $(4,2,1)\oplus(\bar{4},2,1)$ under the global
$SU(4)\times SU(2)\times SU(2)$ symmetry. These are
the symmetries and matter representations of the Pati-Salam GUT. It is interesting that
the minimal mirror model we construct that cancels off gravitational 't Hooft anomalies
for \KD fermions generates a well known chiral gauge theory.~\footnote{Early work connecting
symmetric mass generation in lattice theories to grand unified theories can be found in \cite{You:2014vea}.}

While our discussion has focused on geometries with the topology of the
sphere the connection to Pati-Salam is made in flat space.
In this limit one can replace the general triangulation by a hypercubic
lattice and map the (reduced) lattice \KD field into a (reduced) staggered field \cite{Catterall:2020fep,Catterall:2022jky}.

In this case, and assuming we can indeed gap the mirror sector
as described, we claim the continuum limit of that staggered fermion theory would
correspond to a theory with the global symmetries and matter representations of the Pati-Salam
GUT - a chiral gauge theory. 

\section{Conclusions}

In this paper we have focused on constructing a path integral describing
reduced \KD fermions propagating on a curved background space. This problem is analogous
to the problem of defining a path integral for a gauged chiral fermion. However, the \KD case
possesses one key advantage over the chiral fermion problem
because the operator $\Gamma$ that plays the role of $\gamma_5$ survives intact when the theory is discretized on a triangulation of the space.  This allows us to study the problem in a regulated lattice model with an onsite $U(1)$ symmetry and no fermion
doubling.~\footnote{No doubling in this context means that discretization does not
produce any additional low energy states that were not already present in the
continuum theory.} The lattice \KD operator for
reduced fermions maps the space of even cochains to odd cochains and vice versa. 
On a space with non-zero Euler characteristic
it takes the form of a finite rectangular matrix. To integrate out the fermions then requires
the use of singular value decomposition which yields two unitary matrices that
determine orthonormal bases in these two spaces. If the Euler character
of the triangulation is non-zero there is an ambiguity in the phase of the fermion measure arising
from the determinants of these unitary matrices
which becomes gauge field dependent after the $U(1)$ symmetry associated to $\Gamma$ is gauged. This fermion measure problem for reduced \KD fermions is very similar to
that encountered in defining the path integral for chiral fermions. 

As for chiral fermions
one can add mirror fermions to remove this phase ambiguity.
One then discovers that the global $U(1)$ symmetry
of the mirror model is broken via an anomaly to $Z_{2\chi}$.
This anomaly can be computed exactly
on a finite triangulation contradicting the folklore that only systems with an infinite
number of degrees of freedom can generate anomalies. 
Restricting to the sphere $S^4$ and requiring cancellation of the resulting 't Hooft
anomaly one finds that only multiples of four reduced fermions are allowed. 
In the flat space
continuum limit this fermion content corresponds to multiples of sixteen Weyl spinors
and our result
is in agreement with the cancellation of spin-$Z_4$ anomalies
for Weyl fermions. Indeed, the \KD argument can be generalized to
any even dimension and yields results for anomaly cancellation
for a variety of discrete symmetry \cite{Garcia-Etxebarria:2018ajm,RevModPhys.88.035001}. 
We employ this constraint to identify a set of Yukawa interactions that should be
capable of lifting the mirror states to the cut-off without breaking symmetries or
disturbing the light sector. There is already numerical evidence that suggests that
such symmetric mass generation is possible in such theories.

Remarkably, we show that the light sector of the minimal anomaly free
mirror model, in the continuum flat space limit,
is nothing more than the Pati-Salam GUT model. Rather surprisingly the
search for a non-perturbative
construction of anomaly free reduced \KD theories appears to target a class of anomaly free chiral
models. 
Indeed, the mirror model of reduced \KD fields
we describe can be thought of as providing a non-perturbative definition of the Pati-Salam
model.
Indeed, it would seem that any chiral theory with an $SU(2)\times SU(2)\times G$ 
global symmetry can be handled
in this manner as long as 
$G$ contains an complex, irreducible representation whose dimension is a multiple of four.

Of course to create a true chiral gauge theory
one needs to gauge these non-abelian symmetries.
Gauging the group $G$ in the lattice theory is straightforward and follows the procedure given in eqn.~\ref{u1} and the measure is automatically gauge invariant.
Gauging the internal $SU(2)\times SU(2)$ symmetry is more problematic. In the case of
staggered fermions this symmetry is broken to a $Z_2$ subgroup corresponding to
the shift symmetry $\chi(x)\to \chi(x+\mu)\xi_\mu(x)$ where $\chi$ is the staggered fermion
field and $\xi_\mu(x)=\left(-1\right)^{\sum_{i=\mu+1}^D}x _i$. The presence of this exact global $Z_2$ symmetry is enough to ensure restoration
of the global $SU(2)\times SU(2)$ symmetry in the continuum limit \cite{Golterman:1984cy,vandenDoel:1983mf} and one might hope that gauging the $Z_2$ symmetry would allow for an $SU(2)\times SU(2)$ emergent gauge symmetry. However,
the offsite nature of the shift symmetry makes it difficult to gauge the $Z_2$. Recent work
on Hamiltonian lattice systems with similar shift symmetries may offer a path forward \cite{Cheng:2022sgb,Seiberg:2023cdc}.

\section*{Acknowledgements}
This work was supported by the US Department of Energy (DOE), 
Office of Science, Office of High Energy Physics, 
under Award Number DE-SC0009998. 

\bibliography{mirror.bib}

\nolinenumbers

\end{document}